\documentclass[aps,pra,twocolumn,showpacs,floatfix,superscriptaddress]{revtex4-1}
\usepackage[utf8]{inputenc}
\usepackage[T1]{fontenc}
\usepackage[english]{babel}
\usepackage{float}
\usepackage{color}
\usepackage{graphicx}
\usepackage{indentfirst}
\usepackage{amsmath,amsfonts,amsthm,bm}
\usepackage{siunitx}
\begin{document}
%
\title{Non-collinear antiferromagnetic states in Ru-based Heusler compounds induced by biquadratic coupling}
\author{Eszter Simon}
\email{esimon@phy.bme.hu}
\affiliation{Department of Theoretical Physics, Budapest University of Technology and Economic, Budafoki \'{u}t 8, H-1111 Budapest, Hungary}
\author{Andreas Donges}%
\affiliation{Department of Physics, University of Konstanz, D-78464 Konstanz, Germany}
\author{L\'{a}szl\'{o} Szunyogh}
\affiliation{Department of Theoretical Physics, Budapest University of Technology and Economic, Budafoki \'{u}t 8, H-1111 Budapest, Hungary}
\affiliation{MTA-BME Condensed Matter Research Group, Budapest University of Technology and Economics, Budafoki \'{u}t 8, H-1111 Budapest, Hungary}
\author{Ulrich Nowak}
\affiliation{Department of Physics, University of Konstanz, D-78464 Konstanz, Germany}
\date{\today}
\begin{abstract}
We investigate the magnetic properties of Ru$_{2}$Mn$Z$ ($Z$\,=\,Sn, Sb, Ge, Si) chemically ordered full Heusler compounds for zero as well as finite temperatures. Based on first principles calculations we derive the interatomic isotropic bilinear and biquadratic couplings between Mn atoms from the paramagnetic state. We find frustrated isotropic couplings for all compounds and in case of  $Z$\,=\,Si and Sb a nearest-neighbor biquadratic coupling that favors perpendicular alignment between the Mn spins. By using an extended classical Heisenberg model in combination with spin dynamics simulations we obtain the magnetic equilibrium states. 
From these simulations we conclude that the biquadratic coupling, in combination with the frustrated isotropic interactions, leads to non-collinear magnetic ground states in the Ru$_{2}$MnSi and Ru$_{2}$MnSb compounds.
In particular, for these alloys we find two distinct, non-collinear ground states which are  energetically equivalent and  can be identified as $3-q$ and $4-q$ states on a frustrated fcc lattice.
Investigating the thermal stability of the non-collinear phase we find that in case of Ru$_{2}$MnSi the multiple$-q$ phase undergoes a transition to the single$-q$ phase, while in case of Ru$_{2}$MnSb the corresponding transition is not obtained due to the larger magnitude of the nearest-neighbor biquadratic coupling. 
\end{abstract}
\maketitle
\section{Introduction}
Technological and fundamental interest in the antiferromagnetic (AFM) materials is increasingly growing since materials with novel-type antiferromagnetic structures are possible candidates for a new generation of spintronic devices \cite{Jungwirth2018,Duine2011, RevModPhys.90.015005}. 
Antiferromagnetic materials can complement or even replace ferromagnetic (FM) components in spintronic devices with improved properties due to their enhanced stability against the perturbation via external magnetic fields. Many technologically important effects have already been implemented using an AFM material as the main element of the system, such the ultrafast spin dynamics, magneto-transport, or exchange bias effects \cite{PhysRevB.81.144427, Wadley587, PhysRevB.96.064435, PhysRev.102.1413, NOGUES1999203}.

AFM Heusler alloys can be a possible extension of the class of known antiferromagnetic materials, but relatively few AFM Heusler alloys are known with sufficient high N\'eel temperatures \cite{Hirohata_2017}.
Full Heusler compounds, with the chemical formula $X_{2}YZ$, where $X$, $Y$ are transition metal and $Z$ is a $p$ group element, are mostly ferromagnetic, but it can be transformed from the FM to the AFM state by changing the atomic composition. 
In case of the Ni$_{2}$MnAl alloy for instance, if Mn atoms are also placed on the Al sites, these Mn atoms interact antiferromagnetically with the nearest neighbor (NN) Mn atoms on the original sites and this material become a  compensated antiferromagnet in the fully disordered (B2) state \cite{Acet_JApplPhys2002,PhysRevB.92.054438}, a phase which is called structurally induced antiferromagnetism \cite{Galanakis_ApplPhysLett2011}. 
In Ru$_{2}$Mn$Z$ ($Z =$ Sn, Sb, Ge, Si) alloys the formation of antiferromagnetic order is not structurally induced. 
From earlier theoretical and experimental work it is known that the magnetic ground state of the Ru$_{2}$Mn$Z$ series of Heusler alloys corresponds to the second kind AFM order or simply AF2 \cite{GOTOH1995306,ISHIDA1995140}.
This magnetic state was also confirmed by a recent first principles calculation, in which it was shown, that the chemical disorder significantly reduced the transition temperature of the Ru$_{2}$MnSi system \cite{Sergii-Ru2MnZ}. 
The crystal structure of the ordered Ru$_{2}$Mn$Z$ is $L2_{1}$ type, where the Mn atoms fully occupy one of four interpenetrating fcc subblatices of the $L2_{1}$ structure.
In Ref.~\cite{Sergii-Ru2MnZ} it was reported that the competition between the strong antiferromagnetic next nearest neighbor (NNN)  exchange coupling and the weakly ferromagnetic NN exchange coupling, results in the second type AFM ordering. 
In this AFM state, illustrated in Fig. 1, the magnetic atoms on neighboring (111) planes are coupled antiparallel, while atoms within such a (111) plane couple ferromagnetically, leading to a frustration of the ferromagnetic NN exchange. 
\begin{figure} [ht!]
	\centering
	\includegraphics[width=0.8\columnwidth]{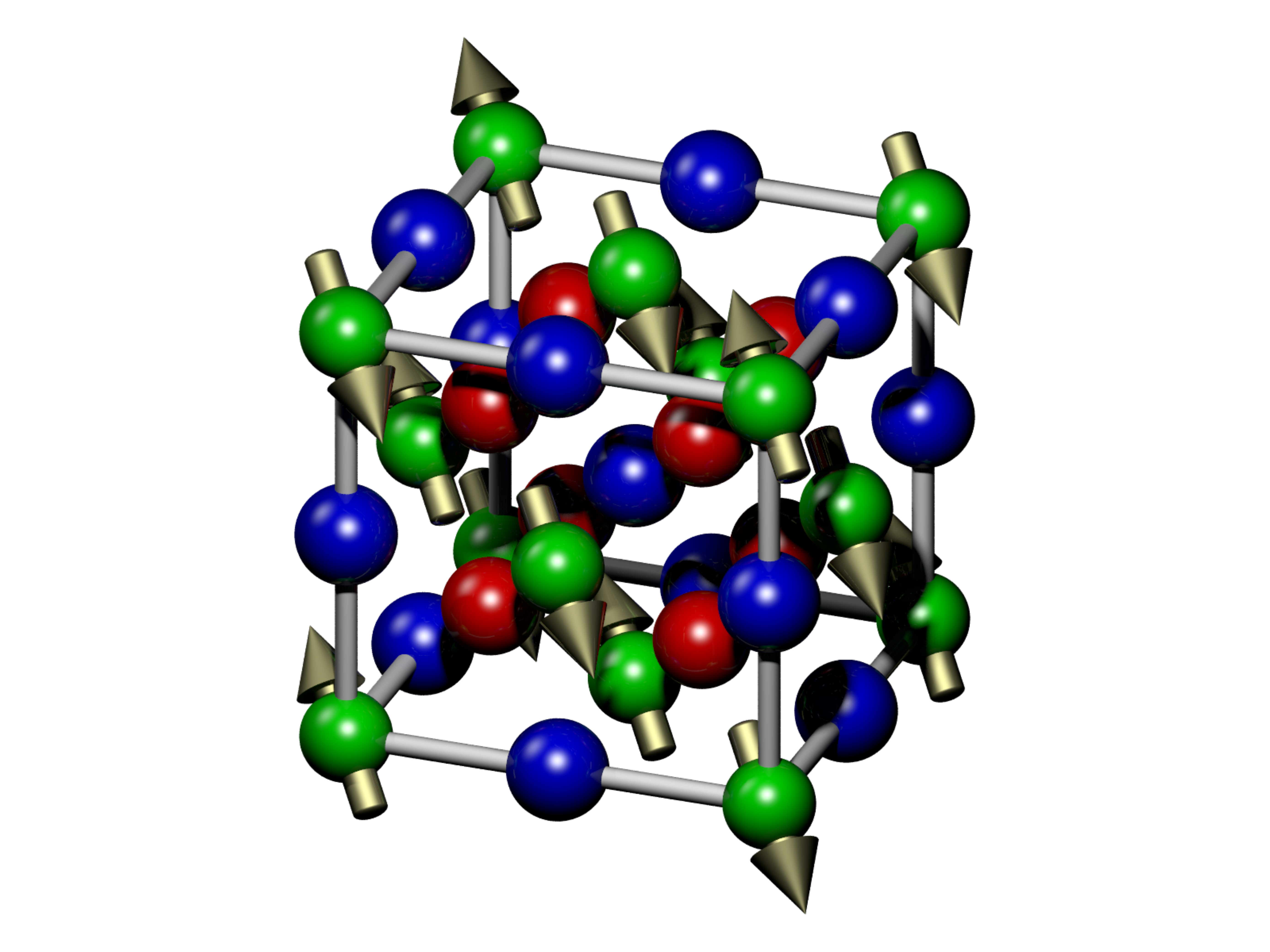}\\
	\caption
	{
		Unit cell of the Ru$_{2}$Mn$Z$ full Heusler compound, with Ru atoms in red, Mn atoms in green, and $Z$ atoms in blue. 
		Golden arrows show the second kind antiferromagnetic order (AF2) on the fcc sublattice of Mn atoms. 
	}
	\label{af2}
\end{figure}

Frustration in the magnetic structure can originate from geometric frustration of a lattice
or from the competition between NN and NNN or even farther exchange interactions. Geometrical frustration of a spin configuration is known for triangular antiferromagnets \cite{Kawamura_1998} or in highly correlated metals \cite{Lacroix_2010}. Frustration with higher-order exchange terms can lead to various exotic magnetic states such as the multiple$-q$ spin order, where the spin texture is characterized by a multiple number of coexisting magnetic modulation $q$ vectors. 
A triple$-q$ state stabilized by four-spin interaction was obtained from model simulations on a triangular lattice \cite{PhysRevLett.101.156402, PhysRevLett.108.096401, PhysRevLett.108.096403,PhysRevB.95.224424} as well as using ab-initio calculations in magnetic thin film systems \cite{PhysRevLett.86.1106} and in itinerant hexagonal magnets \cite{Canepa_2005,Takagieaau3402}. 
In case of the highly correlated Kondo lattice compound UCu$_{5}$ it was shown that a stable $4-q$ state is formed at zero temperature due to the biquadratic coupling \cite{Ueland2012}.

In Ref.~\cite{Sergii-Ru2MnZ} it was mentioned that the biquadratic coupling lifts the degeneracy of the highly frustrated 
AF2 states and positive/negative nearest neighbor biquadratic couplings can lead to collinear/non-collinear magnetic ground states in case of RuMn$_2$(Sn,Ge) and
RuMn$_2$(Sb,Si), respectively.
However, the main focus of Ref.~\cite{Sergii-Ru2MnZ} was on 
the N\'eel temperature of these Heusler alloys which is hardly affected by the weak biquadratic couplings, while the effect of chemical disorder on the spin structure and the N\'eel temperature was investigated in detail using Monte-Carlo simulations.

In this paper we examine the effect of the biquadratic coupling on the magnetic state of the chemically ordered Ru$_{2}$Mn$Z$ Heusler compounds using atomistic spin model simulations. 
The isotropic bilinear and biquadratic couplings are determined from the paramagnetic state using first principles calculations. 
Based on these spin model parameters we employ an extended, classical Heisenberg model and explicitly demonstrate that biquadratic couplings result in a non-collinear magnetic ground state in case of Ru$_{2}$MnSi and Ru$_{2}$MnSb compounds, where the NN biquadratic coupling favors perpendicular alignment between the Mn moments. In these cases we carefully analyse the spin configurations obtained at zero temperature and conclude that $3-q$ and $4-q$ states occur with equal probability as the energy of these states is not dissolved within the applied model.
We also show that, depending on the magnitude of the biquadratic coupling, at finite temperature the non-collinear magnetic phase can transform to the $1-q$ phase.

\section{Calculation details}
We performed self-consistent calculations for the Ru$_{2}$Mn$Z$ compounds within the local spin-density approximation (LSDA) \cite{vosko1980} by using the screened Korringa-Kohn-Rostoker (KKR) method \cite{PhysRevB.49.2721, PhysRevB.52.8807} in the atomic sphere approximation (ASA) by expanding 
the partial waves up to $l_{\rm max} = 3$ ($spdf-$ basis) inside the atomic spheres.
The electronic structure was determined in the paramagnetic state in terms of the scalar relativistic disordered local moment (DLM) scheme \cite{0305-4608-15-6-018}. For all investigated alloys we used 
the experimental lattice constants ($a$) of the $L2_{1}$ lattice structure \cite{Sergii-Ru2MnZ} . To derive the spin model parameters below we employed the spin-cluster expansion technique as combined with the relativistic DLM technique \cite{PhysRevB.83.024401}. 

Our first principles calculations result in a classical spin Hamiltonian of the form 
\begin{equation}
H=-\frac{1}{2}\sum_{i,j}J_{ij} \, \vec{s}_{i}\cdot\vec{s}_{j}-\frac{1}{2}\sum_{i,j} B_{ij}(\vec{s}_{i}\cdot\vec{s}_{j})^{2},
\label{Eq:HamT}
\end{equation}
where $\vec{s}_i$ is the unit vector along the direction of the spin moment of atom $i$. 
In Eq.~(\ref{Eq:HamT}) the first term corresponds to a generalized Heisenberg model where ${J}_{ij}$ is the isotropic exchange interaction. 
The second term describes the isotropic biquadratic interaction between spins $i,\,j$ with the coupling constants $B_{ij}$. 
In the sign convention of Eq.~((\ref{Eq:HamT}), ${J}_{ij} > 0$ describes the ferromagnetic coupling between the magnetic moments, while  ${J}_{ij} < 0$ corresponds to the antiferromagnetic interaction. 
Likewise, $B_{ij}>0$ favors collinear and $B_{ij}<0$ perpendicular alignment of neighboring Mn spins.
These spin model parameters were calculated for $369$ neighbors within a radius of $3a$, where $a$ is the lattice constant of the corresponding system. 

To study the magnetic ground state of the system and the equilibrium state at finite temperatures, we solved the stochastic Landau-Lifshitz-Gilbert (SLLG) equation on a discrete lattice,
\begin{equation}
\frac{\partial \vec{s}_i}{\partial t}= -\frac{\gamma}{(1+\alpha^2)\mu_{\rm s}}\vec{s}_i \times \left(\vec{H}_i + \alpha\,\vec{s}_i \times \vec{H}_i\right),
\label{Eq_SLLG}
\end{equation}
by means of Langevin Dynamics, using a Heun algorithm \cite{lyberatosJPCM93,nowakBook07}. 
The SLLG equation includes the gyromagnetic ratio $\gamma$, a phenomenological damping parameter $\alpha$, and the effective field,
\begin{equation}
 \vec{H}_i = {\vec{\zeta}_i(t)} - \frac{\partial H}{\partial \vec{s}_i } 
=  {\vec{\zeta}_i(t)}   
+\sum_{j(\neq i)}J_{ij}\vec{s}_{j}+2\sum_{j(\neq i)}B_{ij}(\vec{s}_{i}\cdot\vec{s}_{j})\vec{s}_{j} ,
\end{equation}

\noindent
which considers the influence of a temperature $T$ by adding a stochastic noise term $\vec{\zeta}_i(t)$, obeying the properties of white noise \cite{palaciosPRB98},
\begin{equation}
\langle{\vec{\zeta}_i(t)}\rangle=0,
\end{equation}
\begin{equation}
\langle\zeta_{i}^{\eta}(t)\zeta_{j}^{\theta}(t')\rangle=\frac{2k_{B}T\alpha\mu_{s}}{\gamma}\delta_{ij}\delta_{\eta\theta}\delta(t-t').
\end{equation}
Here $i,\;j$ denote lattice sites and $\eta$ and $\theta$ Cartesian components of the stochastic noise.

We used a $884736$ atoms in the unit cell of the Mn sublattice with periodic boundary conditions and the couplings included up to the 6th NN shell.
Two kinds of simulations were performed:
To check the magnetic ground state we cooled the system down slowly, starting from the paramagnetic state using $\alpha=0.0001$.
For calculating the specific heat, this parameter was set to $0.5$, and the system was heated up incrementally, starting from the previously determined ground state spin configuration.

\section{Results}
\begin{figure} [t!]
\centering
\includegraphics[width=1.00\columnwidth]{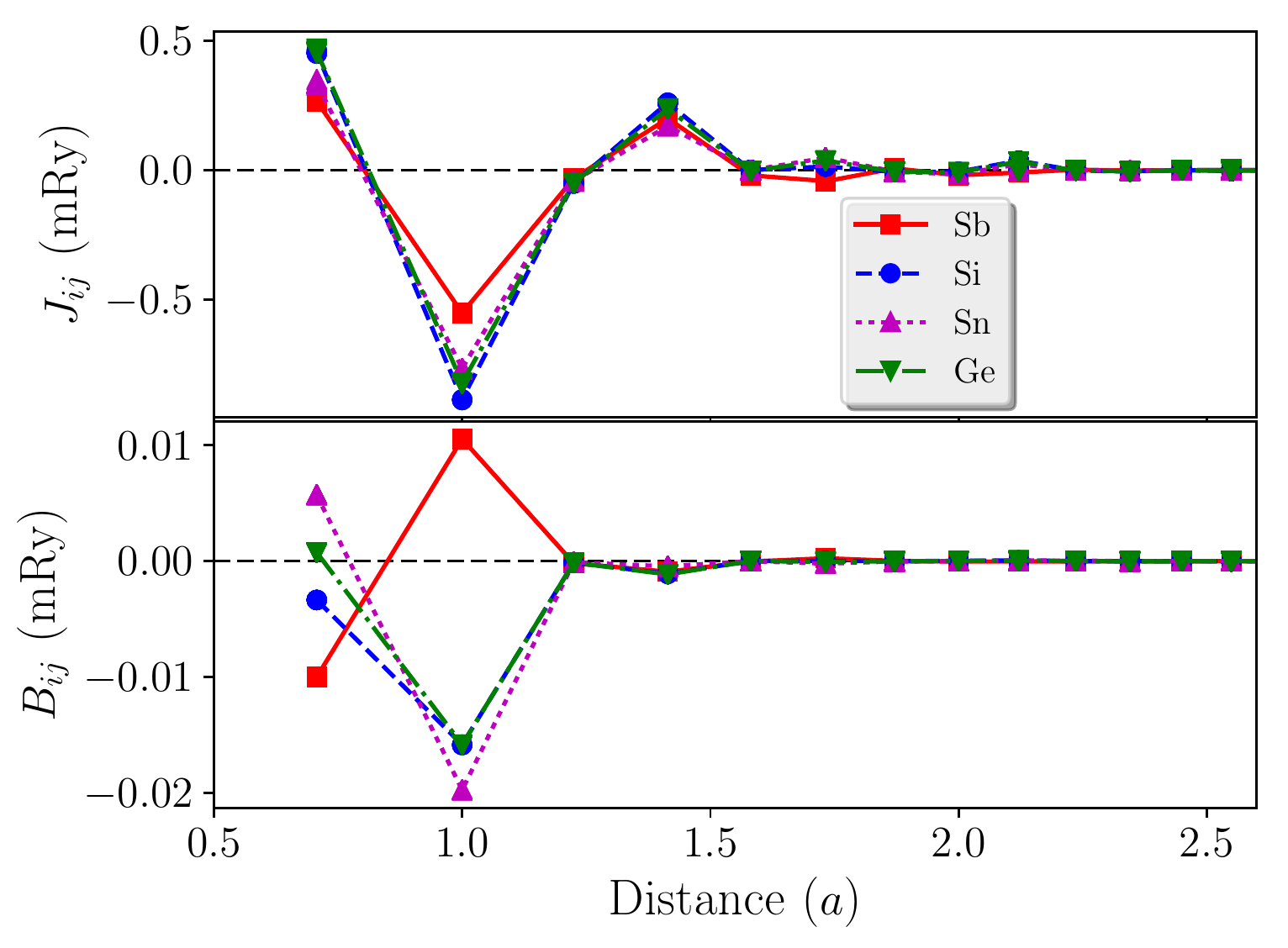}\\
\caption{Calculated Mn-Mn exchange interactions, $J_{ij}$  and biquadratic couplings, $B_{ij}$ for  Ru$_{2}$Mn$Z$ ($Z=$Sn, Si, Sb, Ge) alloys as a function of distance between Mn atoms in unit of the lattice constant, $a$. }
\label{fig_jij}
\end{figure}
\subsection{Spin model parameters}
First we performed first principles electronic structure calculations for the Ru$_{2}$Mn$Z$ series of the full Heusler compounds in the $L2_{1}$ geometry, where the electronic structure was determined in the paramagnetic state. 
Similarly as in Ref.  \cite{Sergii-Ru2MnZ}, we also found that the smallest local magnetic moment of Mn has the Ru$_{2}$MnSi system with value of $2.90 \,\mu_{B}$, while for Ru$_{2}$MnSb the largest Mn magnetic moment were obtained with value of $3.55 \, \mu_{B}$. 
From the self consistent potentials we determined the isotropic bilinear and biquadratic couplings between the Mn atoms for the Ru$_{2}$Mn$Z$ compounds employing the spin-cluster expansion technique. 
Note, that according to the definition of the spin Hamiltonian in Eq.~(\ref{Eq:HamT}), our spin-model parameters are twice as large than in Ref.~\cite{Sergii-Ru2MnZ}.
The NN isotropic exchange coupling is ferromagnetic for all cases and smaller than the second NN interaction as Fig. \ref{fig_jij}(a) shows. 
The second and third NN couplings are antiferromagnetic, while the fourth NN interaction is ferromagnetic again and the magnitude of further NN couplings become negligible beyond the fourth-NN shell.

In Fig \ref{fig_jij}(b) the biquadratic couplings are also presented as a function of the distance between the Mn atoms. 
The NN biquadratic coupling is positive for Ru$_{2}$MnGe and Ru$_{2}$MnSn, while negative for Ru$_{2}$MnSi and Ru$_{2}$MnSb; and beyond the second NN biquadratic coupling, the $B_{ij}$ decays rapidly. 
According to the sign convention in Eq.~(\ref{Eq:HamT}), negative $B_{ij}$ means that the favored configuration between the Mn moments is perpendicular. 

\subsection{Magnetic ground state}
In agreement with previous theoretical results \cite{ISHIDA1995140,Sergii-Ru2MnZ}, the ferromagnetic NN and antiferromagnetic NNN exchange interactions presented in Fig.~\ref{fig_jij} strongly indicate an AF2 magnetic ground state \cite{moran1994} for all investigated alloys. As discussed in Ref.~\cite{Herrmann_Ronzaud_1978} there are several equivalent AF2 structures that are linear combinations of the collinear AF2 states depicted in Fig.~\ref{af2} related to  the four possible wave vectors, $\vec{q}_1=\frac{\pi}{a}(1,1,1)$, $\vec{q}_2=\frac{\pi}{a}(-1,1,1)$, $\vec{q}_3=\frac{\pi}{a}(1,-1,1)$ and $\vec{q}_4=\frac{\pi}{a}(1,1,-1)$, $a$ being the lattice constant of the fcc lattice. Such spin configurations can be described as multiple$-q$ states,
\begin{equation}
\vec{s}_i = \frac{1}{\sqrt{M}} \sum_{n=1}^M \vec{s}^{M}_n \exp{ \left( i \vec{q}^M_n \vec{R}_i \right) } \,
\end{equation}
where $M \in \{ 1,2,3,4 \}$, $\vec{q}^M_n \in \{ \vec{q}_1, \vec{q}_2, \vec{q}_3, \vec{q}_4 \}$ and $\vec{R}_i$ stands for the lattice vector of site $i$. The generating spin vectors $\vec{s}^{M}_n$ of unit length should be determined such that $\vec{s}_i$ will also be unit vectors. From this condition it can easily be deduced that a $2-q$ structure ($M=2$) is described by four magnetic sublattices associated with four neighboring sites of the fcc lattice forming a regular tetrahedron, with spins being aligned either antiparallel or normal to each other. Both the $3-q$  ($M=3$) and the  $4-q$  ($M=4$) textures can be described by doubling this magnetic unit cell, thus by eight magnetic sublattices. In one tetrahedron the spin vectors make an angle of either $\SI{109.5}{\degree}$ (tetrahedral angle) or $\SI{70.5}{\degree}$  ($=\SI{180}{\degree}-\SI{109.5}{\degree}$), i.e. $\vec{s}_i\cdot \vec{s}_j = \pm1/3$ ($i \ne j$), while the spins in the other tetrahedron are reversed as compared to the first one. One important difference between these two states is that for a $3-q$ state one can always find a (111) type of plane which is magnetically compensated, while in a $4-q$ state the spins in each tetrahedra compensate each other, thus, up to a global rotation of the spin vectors, the $4-q$ state restores the cubic symmetry  \cite{Herrmann_Ronzaud_1978}.

An effective isotropic spin model for the eight magnetic sublattices can easily be set up by making use of the cubic symmetry of the underlying fcc lattice. Let us denote the sublattices corresponding to the two tetrahedral unit cells by $a$, $b$, $c$, $d$ and $A$, $B$, $C$, $D$, where the sites in the sublattices labeled by the same small and capital letter are shifted by $a \, (n, m, k)$ with $n,m,k$ being integers. The corresponding bilinear spin model,
\begin{equation}
H_{\rm bl} = -\frac{1}{2} \sum_{\alpha,\beta} 
J_{\alpha \beta} \, {\vec s}_\alpha \cdot {\vec s}_\beta \, ,
\end{equation}
$\alpha$ and $\beta$ labelling sublattices, contains only three independent  parameters: $J_{\rm aa}$, $J_{\rm aA}$ and $J_{\rm ab}$ that refer to the effective intrasublattice interaction, to the interaction between the same kind of sublattices in the two tetrahedra and to the interaction between different kinds of sublattices, respectively. In fourth nearest neighbor approach, these interactions can be expressed as 
\begin{equation}
J_{\rm aa}=6 J_4, \; J_{\rm aA}=6 J_2, \;  J_{\rm ab}=2 J_1 + 4 J_3\, ,
\label{Jsubl}
\end{equation}
where $J_n$ denotes the atomic exchange parameters between the $n^{\rm th}$ neighbors. It is then simple to show that in the billinear model all the multiple$-q$ states discussed above have the same energy, $E_{\rm AF2}=\frac{1}{2} (J_{\rm aA}-J_{\rm aa})$ as normalized to one magnetic atom. 

By including biquadratic couplings into the sublattice spin model,
\begin{equation}
H_{\rm bq} = -\frac{1}{2} \sum_{\alpha,\beta} 
B_{\alpha \beta} \, ( {\vec s}_\alpha \cdot {\vec s}_\beta )^2 \, ,
\end{equation}
the coupling coefficients $B_{\alpha \beta}$ have the same structure as mentioned in context of Eq.~\eqref{Jsubl}. The biquadratic interactions change the energy of the AF2 states by 
\begin{align}
E_{\rm bq}^{1-q} & = -\frac{1}{2} ( B_{\rm aa} + B_{\rm aA} + 6 B_{\rm ab} ) 
\label{ebq:1-q} \\
 E_{\rm bq}^{2-q} & = -\frac{1}{2} ( B_{\rm aa} + B_{\rm aA} + 2 B_{\rm ab} )
 \label{ebq:2-q}
 \end{align}
 and
 \begin{align}
E_{\rm bq}^{3-q} & =  E_{\rm bq}^{4-q} = -\frac{1}{2} ( B_{\rm aa} + B_{\rm aA} + \frac{2}{3} B_{\rm ab} ) \, .
\label{ebq:34-q}
\end{align}
This means that the biquadratic coupling between different kinds of sublattices, $B_{\rm ab}$, lifts the degeneracy of the AF2 states: if $B_{\rm ab}>0$ then the collinear $1-q$ states will be the ground state and for $B_{\rm ab}<0$ the non-collinear $3-q$ and $4-q$ states have the lowest energy. The $2-q$ state will always be of higher energy than either the $3-q$, $4-q$ or the $1-q$ state, and is therefore not expected to be observed. It should be noted that it is primarily the NN biquadratic coupling $B_1$ which makes a difference in the energy of the AF2 states, since $B_{\rm ab} \sim 2 B_1$, while the NNN biquadratic coupling $B_2$ is irrelevant in lifting the degeneracy.

From the calculated spin-model parameters we determined the magnetic ground state of the chemically ordered Ru$_{2}$Mn$Z$ Heusler compound via simulated annealing. In all cases, the obtained configurations are characterized by alternating (111) planes with spins of reversed directions, stemming from the strong antiferromagnetic NNN exchange interaction. 
According to the considerations above, we found that depending on the sign of  $B_{1}$ different antiferromagnetic magnetic ground states were formed.
As shown in Fig.~\ref{fig_jij} in case of the Ru$_{2}$MnGe and Ru$_{2}$MnSn compounds, the nearest-neighbor biquadratic coupling, $B_\text{1}$ is positive, while for the Ru$_{2}$MnSi and Ru$_{2}$MnSb alloys it is negative. 
For $B_\text{1} > 0$, i.e. for $Z=$ Sn and Ge, all spins are coupled ferromagnetically on each (111) plane, leading to an energetically favored collinear $1-q$ state, as it is also observed in other fcc-AFMs such as the metal oxides $M$O ($M$\,=\,Mn, Fe, Co, Ni) for instance \cite{lit:PR-Roth-MO}. 
In case of $B_\text{1} < 0$, i.e. for $Z=$ Si and Sb, the magnetic order in the (111) plane is more complicated, as neighboring spins want to align perpendicular to each other.

\begin{figure} [t!]
\centering
\includegraphics[width=1.00\columnwidth]{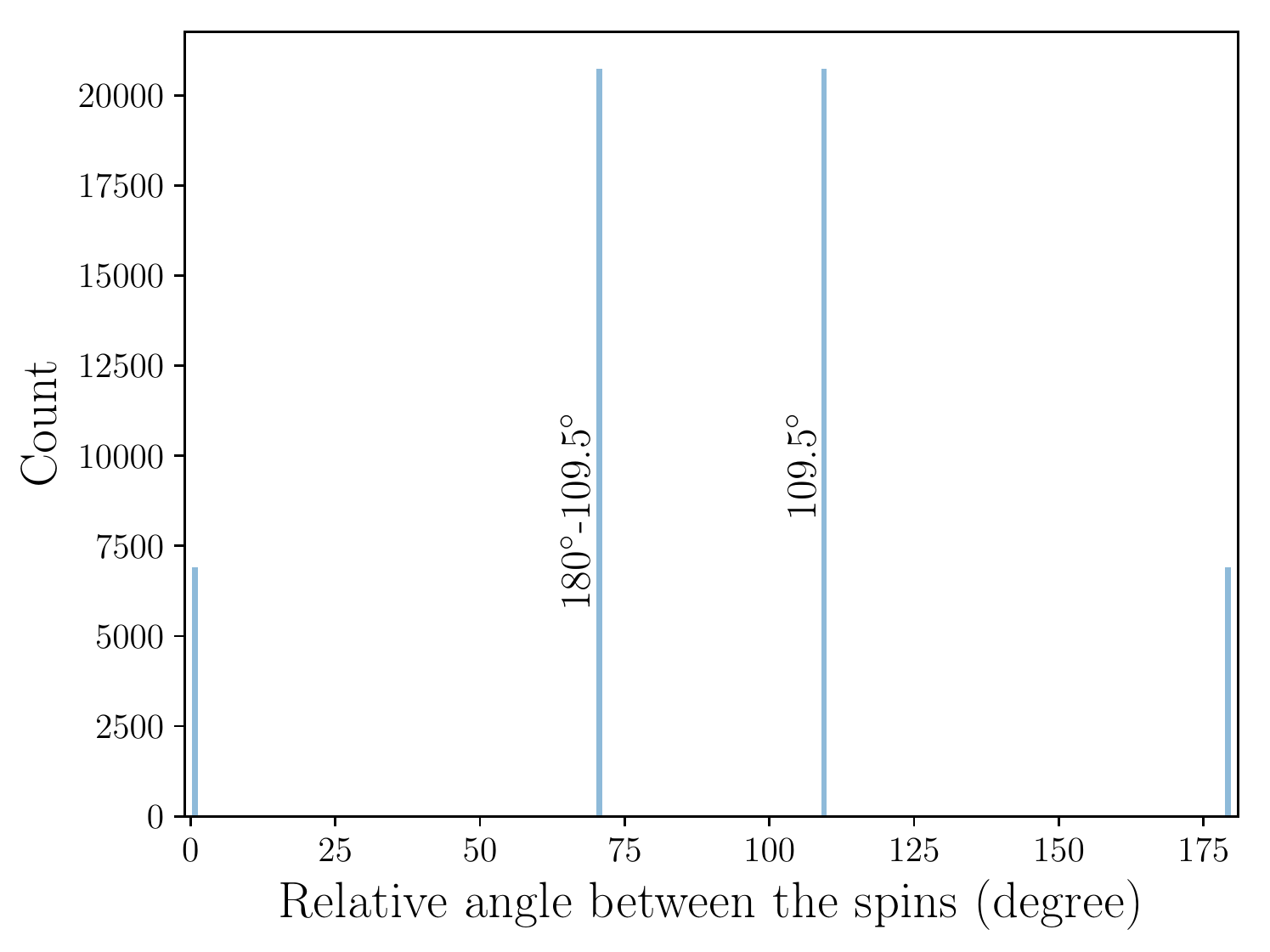}\\
\caption{Distribution of the angles between an arbitrarily chosen spin $\vec s_i$ and the remaining lattice spins $\vec s_{j\neq i}$, in case of negative nearest-neighbor biquadratic coupling. Beyond the usual $0^{\circ}$ and $180^{\circ}$ angles, other tetrahedral, $109.5^{\circ}$ and $180^{\circ}-109.5{\circ}$ angles appear as well due to the biquadratic coupling.}
\label{fig_deg}
\end{figure}

\begin{figure}[htb!]
\includegraphics[width=0.95\columnwidth]{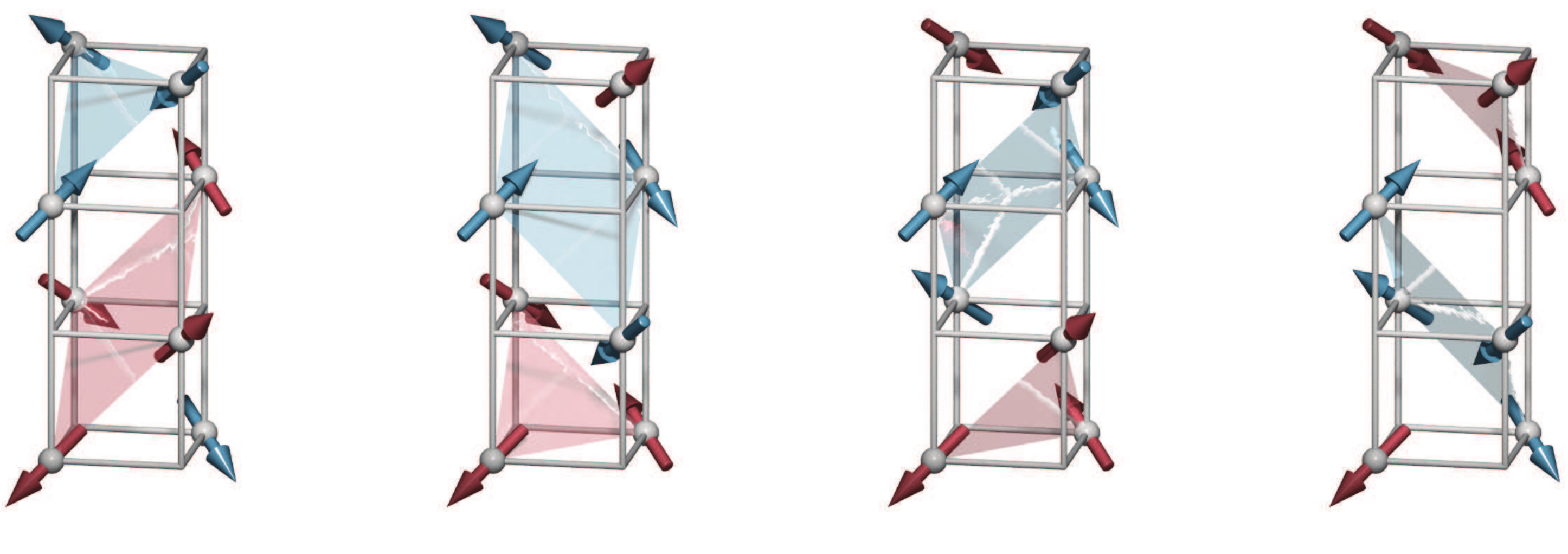} \\
\includegraphics[width=0.95\columnwidth]{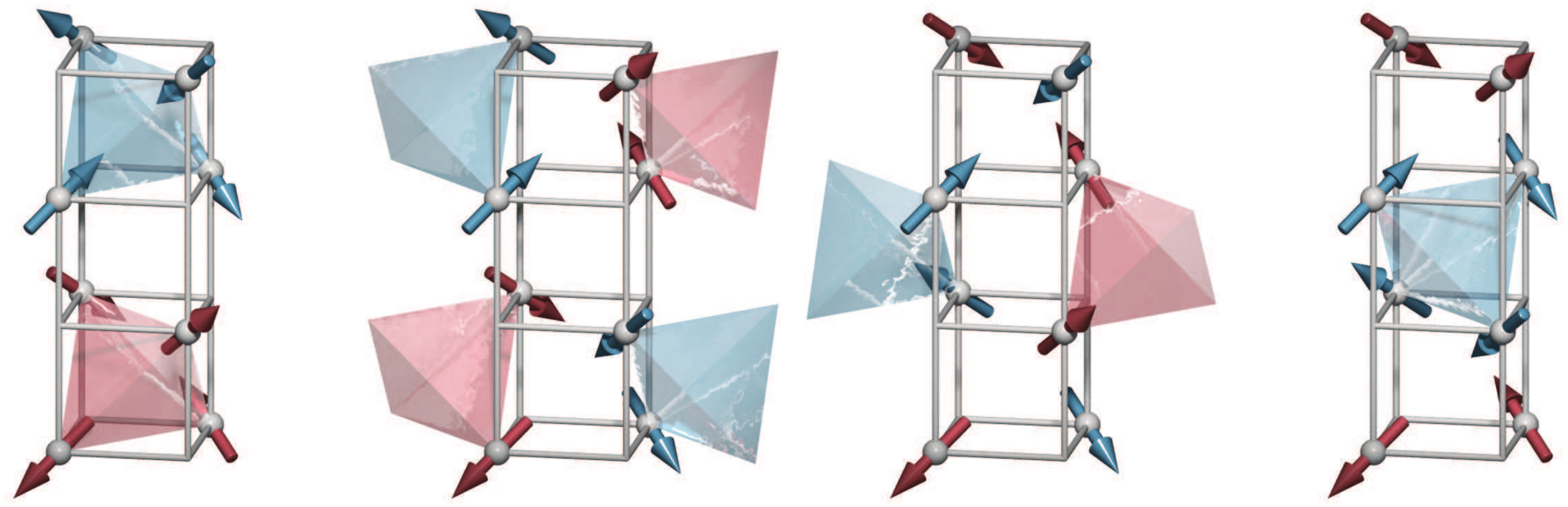} \\
\caption{Magnetic ground state configurations simulated for  Ru$_{2}$Mn$Z$ ($Z$\,=\,Si,Sb) Heusler compounds with negative NN biquadratic coupling, $B_\text{1}$. The upper panel shows different variations of $3-q$ states, where the fully compensated (111) planes are indicated with red and blue colors.
In the lower panel different variations of $4-q$ states can be seen, with fully compensated tetrahedra indicated by red and blue colors.}
\label{fig_gs}
\end{figure}

We calculated the distribution of the relative angles between the spin moments in the simulated cell of spin configuration and found that, beyond $0^{\circ}$ and  $180^{\circ}$, relative angles of \SI{109.5}{\degree} and \SI{70.5}{\degree} occur  in the system as demonstrated in Fig.~\ref{fig_deg}. 
This distribution obviously corresponds to the $3-q$ and $4-q$ states displayed  in the upper and lower panels of Fig.~\ref{fig_gs}, respectively.
From these figures it can be inferred that by fixing an arbitrary spin in a magnetic unit cell, there is one spin aligned antiparallel to this spin, while three spins make  \SI{109.5}{\degree} and three spins make \SI{70.5}{\degree} with the chosen spin. 
This explains the ratios of 1:3:3:1 for the occurrence of the four relative angles  in Fig.~\ref{fig_deg}. 

For the case of $B_1 < 0$ we determined the probability of the $3-q$ and $4-q$ states in the ground state by relaxing a total of 512 random configurations of $48^3$ spins. We detected $3-q$ states in 253 cases (\SI{49.4}{\percent}), while $4-q$ states occured in 259 cases (\SI{50.6}{\percent}). These simulational results indicate that the magnetic ground state of the system corresponds to an ensemble of $3-q$ and $4-q$ states 
of equal probability.

We also calculated the difference between the energy of the ground state 
configurations and that of the collinear $1-q$ state and obtained $\Delta E_\text{Si} = -0.0096$\,mRy/spin and $\Delta E_\text{Sb} = -0.0275$\,mRy/spin, respectively. These values are in good agreement with the analytical expressions in Eqs.~\eqref{ebq:1-q} and \eqref{ebq:34-q}, implying $E_{\rm bl}^{3-q} -  E_{\rm bl}^{1-q} =  \frac{8}{3} B_{ab} \sim \frac{16}{3} B_{1}$,
which gives $-0.0089$\,mRy/spin for Si and $-0.0266$\,mRy/spin for Sb.

\subsection{Finite temperature simulations}
\begin{figure} [ht!]
\centering
\includegraphics[width=1.0\columnwidth]{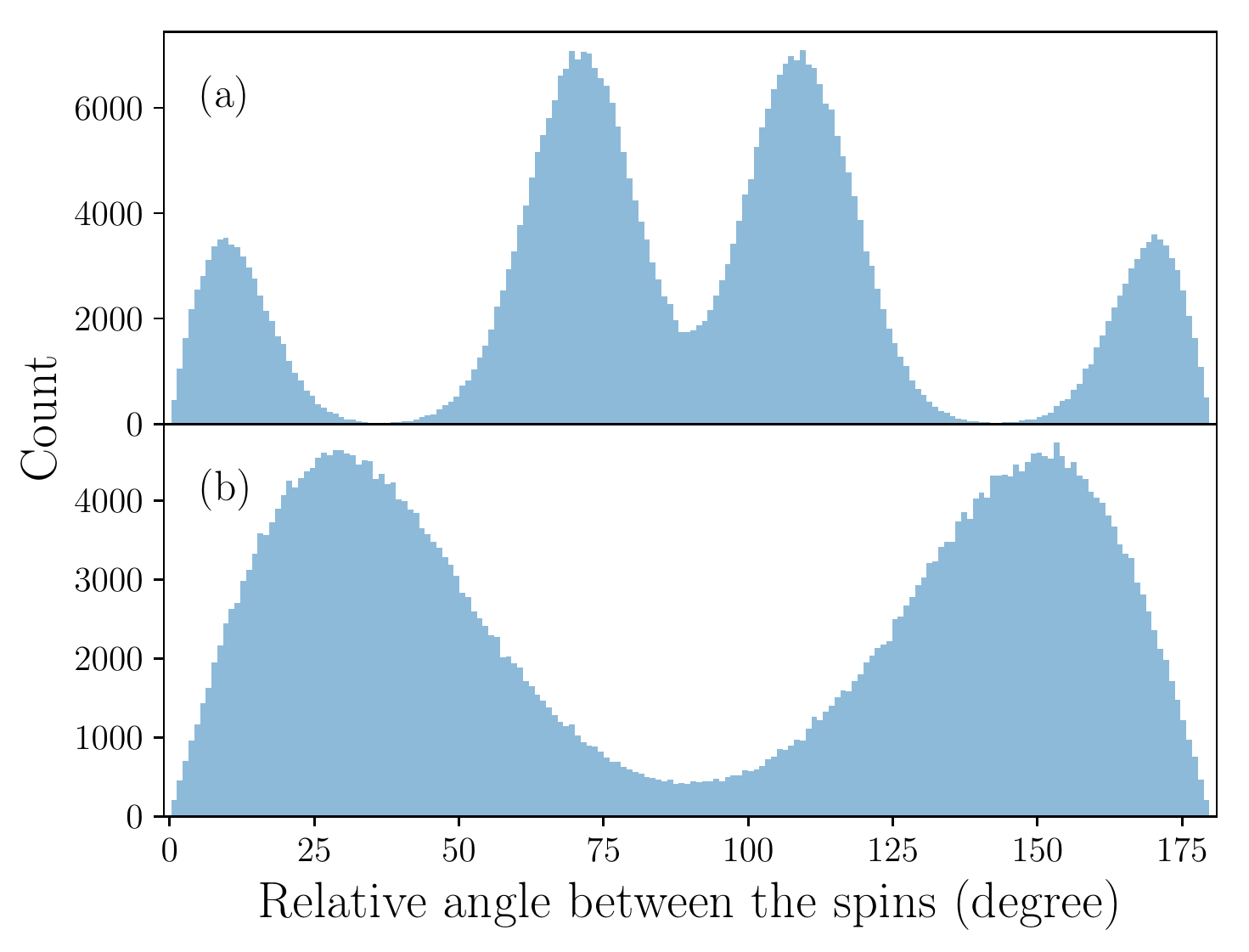}
\caption{Distribution of the angle between an arbitrarily chosen spin $\vec s_i$ and the remaining lattice spins $\vec s_{j\neq i}$ for the Ru$_{2}$MnSi system at two different temperatures. At $T=20$K (a) the $4-q$ state is still stable, while at $T=214$K (b) antiferromagnetic $1-q$ state is formed.}
\label{fig_conf_temp}
\end{figure}

Next, we investigate the equilibrium phases of the Ru$_{2}$MnSi and Ru$_{2}$MnSb compounds at finite temperatures. 
As it was demonstrated in Fig.~\ref{fig_deg} in terms of relative angles, the magnetic ground state of these systems is non-collinear due to the negative NN biquadratic coupling. For the case of Ru$_{2}$MnSi,
the distribution of the relative angles between the spins at two temperatures is shown in Fig.~\ref{fig_conf_temp}. 
From Fig.~\ref{fig_conf_temp}(a) it is obvious that at $T=20$\,K the non-collinear state remains stable  as the distribution of the relative angles shows a four-peak structure. At $214$\,K, however, only two peaks appear in the distribution, which indicates a magnetic phase transition to the $1-q$ phase at higher temperatures.
 
\begin{figure} [ht!]
\centering
\includegraphics[width=1.00\columnwidth]{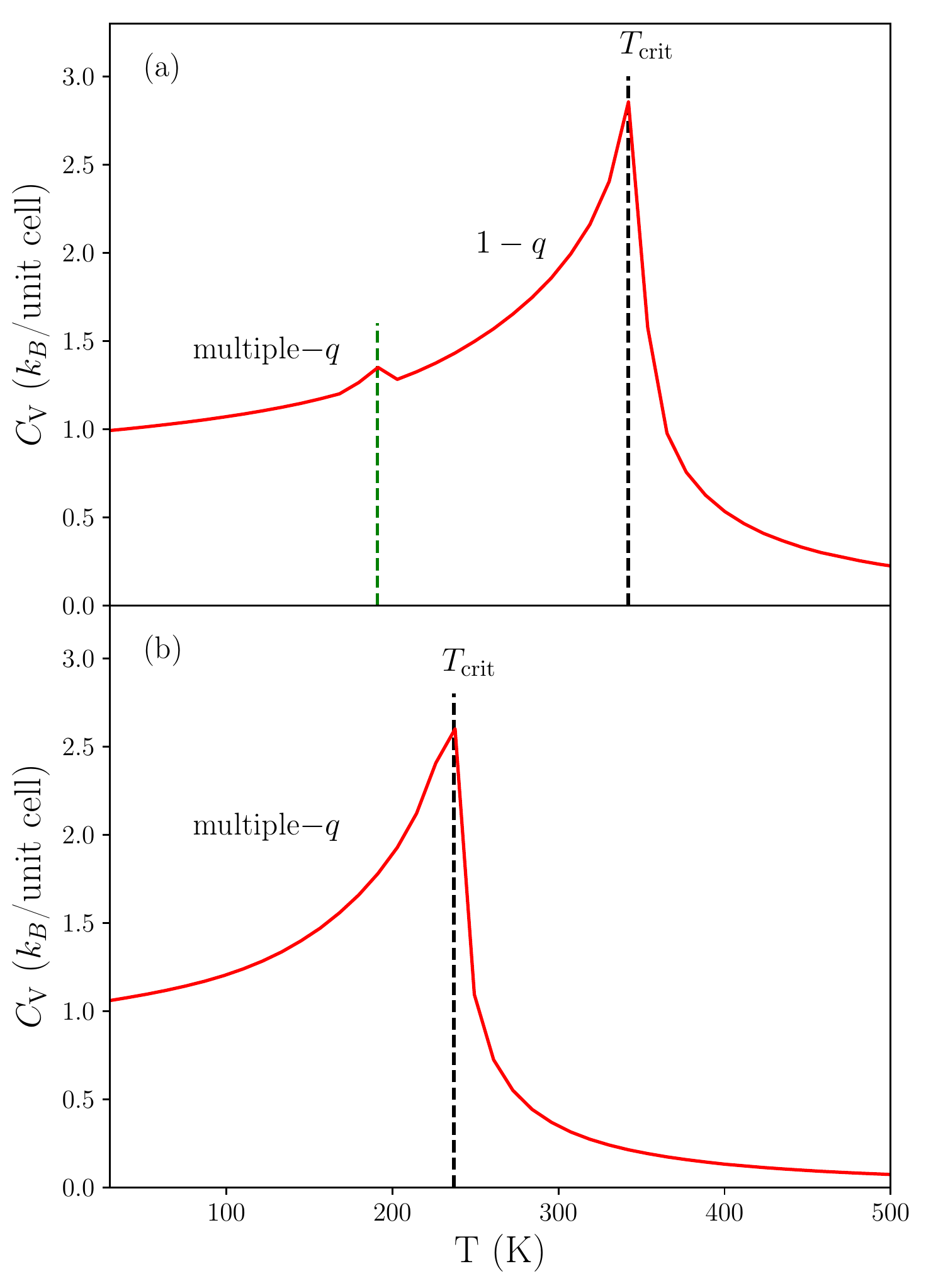}\\
\caption{Calculated heat capacity as function of temperature for the chemically ordered full Heusler compounds Ru$_{2}$MnSi (a) and Ru$_{2}$MnSb (b). The dashed lines denote the phase transitions. In case of Ru$_{2}$MnSi the multiple$-q$ state transformes into the antiferromagnetic $1-q$ state.}
\label{fig_cv}
\end{figure}

To identify the different magnetic phases as a function of the temperature, we determined the heat capacity of the Ru$_{2}$MnSi and Ru$_{2}$MnSb Heusler compounds which we present in Fig. \ref{fig_cv}. 
In case of the Ru$_{2}$MnSi system the heat capacity shows two phase transitions (see Fig. \ref{fig_cv}(a)). The low-temperature phase below $200$\,K possesses the non-collinear state while the higher-temperature phase corresponds to the $1-q$ state, with a critical temperature of $T_{\rm crit}=342$ K above which the system is paramagnetic. 
For the Ru$_{2}$MnSb system only one phase transition is observed from the heat capacity (see Fig. \ref{fig_cv}(b)), where the system transforms directly from the non-collinear state into the paramagnetic one. For this alloy the NN biquadratic coupling is two times larger than in case of Ru$_{2}$MnSi, which explains the absence of the 
multiple$-q$ to $1-q$ phase transition.  
For both systems the critical temperature differs somewhat from the previously reported values in Ref. \cite{Sergii-Ru2MnZ} due to the different NN isotropic couplings but the obtained critical temperatures are also in good agreement with the experiment \cite{KANOMATA20061}, see Table ~\ref{table1}.
\begin{table}[ht!]
\centering
\begin{ruledtabular}
\begin{tabular}{c c c c}
                                                & SD sim. & MC sim. &  exp. \\
                                                &  this work & Ref. \cite{Sergii-Ru2MnZ} &  Ref. \cite{KANOMATA20061} \\
\hline
Ru$_{2}$MnSb &        238                &                 180                            &      195        \\
Ru$_{2}$MnSi &         342               &                   415                           &       313      \\
\end{tabular}
\end{ruledtabular}
\caption{Critical temperatures  for Ru$_{2}$MnSb and Ru$_{2}$MnSi alloys in the present work from spin dynamics simulations and in Ref. \cite{Sergii-Ru2MnZ} from Monte-Carlo simulations. The experimental transition temperatures are also presentes as reported in Ref. \cite{KANOMATA20061}.}
\label{table1}
\end{table}

\section{Conclusions}
In conclusion, we examined the magnetic equilibrium states of the Ru$_{2}$Mn$Z$ ($Z$\,=\,Sn, Sb, Ge, Si) chemically ordered full Heusler alloys at zero and finite temperatures within a multi-scale simulation approach.
We performed first principles calculations in the paramagnetic state via the scheme of disordered local moments and obtained the isotropic and biquadratic couplings between the Mn atoms using the spin-cluster expansion technique. 
We found frustrated bilinear couplings for all considered systems and nearest neighbor biquadratic coupling that favors non-collinear alignment between the Mn atoms in case of Ru$_{2}$MnSi and Ru$_{2}$MnSb compounds, while collinear alignment in case of Ru$_{2}$MnGe and Ru$_{2}$MnSn. 
The frustrated isotropic interactions with the biquadratic coupling lead to a non-collinear antiferromagnetic state in case of  Ru$_{2}$MnSi and Ru$_{2}$MnSb alloys. This non-collinear magnetic ground state comprises energetically equivalent $3-q$ or $4-q$ states that occur with equal probability.

We investigated the thermal stability of the non-collinear state and found a transition to the collinear antiferromagnetic order ($1-q$ state) around $200$K for the Ru$_{2}$MnSi alloy. 
In case of Ru$_{2}$MnSb the nearest-neighbor biquadratic coupling was two times larger than in case of Ru$_{2}$MnSi, thus, the non-collinear state was stable against thermal fluctuations and not transformed to the $1-q$ state below the paramagnetic phase transition. 

From powder neutron diffraction experiments a spin-reorientation transition from [110] to [111] direction was inferred at about 100~K \cite{GOTOH1995306} for Ru$_2$MnSb and a corresponding peak in the magnetization curve of Ru$_2$MnSb was also detected in Ref.~\cite{KANOMATA20061}.  In the same work a pronounced peak in the magnetization was found at 25~K for Ru$_2$MnSi, but its origin was not clearly understood. Although the existence of the multiple$-q$ states has not yet been confirmed experimentally in the investigated Heusler alloys,
our results indicate the importance of higher order exchange interactions in the magnetic equilibrium state of the antiferromagnetic materials.
We expect that the present work may motivate further experimental investigations of antiferromagnetic Heusler alloys and provide a possible application of the non-collinear antiferromagnetic materials. 

\begin{acknowledgments}
This work was supported by the National Research, Development and Innovation Office of Hungary under Projects No. PD120917 and No. K115575, as well as by the BME Nanotechnology FIKP grant (BME FIKP-NAT).
The authors would like acknowledge NIIF for awarding us access to resource based in Hungary at Debrecen.
\end{acknowledgments}
\bibliographystyle{apsrev4-1}
\bibliography{references}
\end{document}